\documentclass[reprint,superscriptaddress,amsmath,amssymb,aps,prl]{revtex4-2}

\usepackage{graphicx,dcolumn,bm,physics,soul,xcolor}
\usepackage[colorlinks=true, linkcolor=black, citecolor=black, urlcolor=black]{hyperref}

\newcommand{\aibte}{\textit{ai}BTE}
\newcommand{\bb}{\textcolor{black}}
\newcommand{\vv}{\textcolor{black}}

\def\oden{Oden Institute for Computational Engineering and Sciences, The University of Texas at Austin, 201 E. 24$^{th}$ Street, Austin, TX 78712, USA}
\def\utphysics{Department of Physics, The University of Texas at Austin, Austin, TX 78712, USA}
\def\ucphysics{Department of Physics, University of California, Berkeley, Berkeley, CA, USA 94720}
\def\lbl{Materials Sciences Division, Lawrence Berkeley National Laboratory, Berkeley, CA, USA 94720}
\def\usc{Mork Family Department of Chemical Engineering and Materials Science, University of Southern California, Los Angeles, CA, USA 90089}

\begin{document}

\title{First-principles predictions of carrier mobility with record accuracy\\ using GW perturbation theory}

\author{Nick Pant}
\affiliation{\oden} 
\affiliation{\utphysics} 

\author{Sabyasachi Tiwari}
\affiliation{\oden} 
\affiliation{\utphysics} 

\author{Steven G. Louie}
\affiliation{\ucphysics}
\affiliation{\lbl} 

\author{Zhenglu Li}
\affiliation{\usc} 

\author{Feliciano Giustino}
\email{fgiustino@oden.utexas.edu}
\affiliation{\oden} 
\affiliation{\utphysics} 

\date{\today}

\begin{abstract}
Accurate prediction of carrier mobility is critical for the discovery and design of next-generation electronic materials. Despite sustained progress, state-of-the-art \textit{ab initio} methods remain limited by the approximate treatment of electron-phonon interactions at the density functional theory level. Here, we demonstrate that incorporating many-body GW corrections to both the electronic band structure and electron-phonon couplings when solving the \textit{ab initio} Boltzmann transport equation yields a mean absolute relative error of just 11\% for electron mobilities across benchmark semiconductors, including Si, GaAs, GaP, diamond, and SiC\vv{. The common practice of neglecting GW corrections to the electron–phonon interaction can lead to mobility errors exceeding 50\%.} The present findings highlight the importance of many-body \bb{GW self-energy} effects in carrier transport simulations, and provides fundamental insights into how many-body electron--phonon interactions govern charge transport in crystalline solids.
\end{abstract}

\maketitle

The thermal motion of atomic nuclei in solids gives rise to an aperiodic time-dependent potential that scatters electrons. This process, known as electron-phonon scattering \cite{giustino2017electron}, imposes an intrinsic limit on carrier mobility in semiconductors. A microscopic theory of charge transport that can predict mobility directly from atomic structure is essential for discovering materials for next-generation electronics and quantum technologies~\cite{jena2022quantum, SRCDecadalPlan2021, liu2021promises, claes2025phonon}. 

The state-of-the-art approach for predicting carrier mobility is the \textit{ab initio} Boltzmann transport equation (\aibte), which uses electronic band structures and electron-phonon couplings computed within density functional theory (DFT)~\cite{ponce2020first}. Despite substantial advances in computational methods and software over the past two decades~\cite{lee2023electron, zhou2021perturbo, cepellotti2022phoebe, gonze2020abinit}, DFT-based mobility predictions may exceed experimental values by 100\% or more in several materials classes \cite{ponce2021first}. Incorporating many-body quasiparticle corrections to the band energies via the GW approximation \cite{hybertsen1986electron, onida2002electronic} can improve agreement in some cases \cite{ma2018first, ponce2018towards, ponce2019route}, but in others, it leads to even larger deviations. These discrepancies persist even when comparing to measurements in undoped, high-purity samples, indicating that they must originate from intrinsic limitations of the theory rather than extrinsic sample quality.

Here, we show that the existing limitations of DFT-based transport calculations can largely be overcome by describing not only band structures, but also electron-phonon couplings, using many-body Green's functions methods beyond DFT. To this end, we combine a state of the art \aibte\ framework \cite{lee2023electron} with the recently developed GW perturbation theory (GWPT), which incorporates the effects of the electronic self-energy into the electron-phonon matrix elements, \cite{li2019electron} for many-body calculations of electron-phonon interactions~\cite{lazzeri2008impact, gruneis2009phonon, faber2011electron, yin2013correlation, mandal2014strong, monserrat2016correlation, faber2015exploring}. By performing calculations for prototypical semiconductors, we find that GW self-energy effects modify electron-phonon coupling matrix elements by up to 80\% by correcting the DFT overscreening of the scattering potential. Upon including this effect, we obtain mobilities on average within 11\% of experimental measurements on high-purity single crystal samples of Si, GaAs, GaP, diamond, and 3C-SiC, marking a record accuracy for fully \textit{ab initio} transport calculations.

\begin{figure*}
    \includegraphics{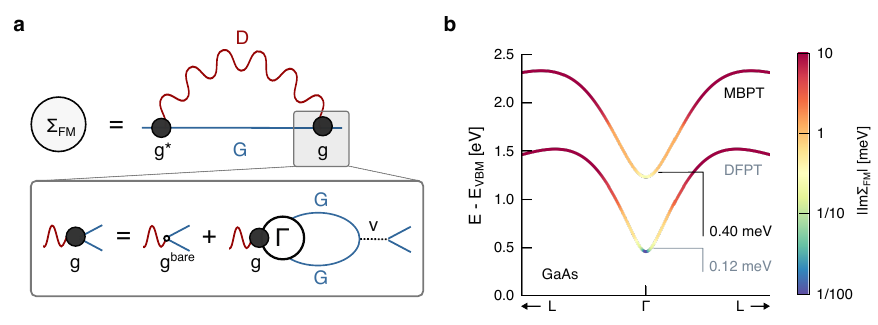}
    \caption{(a) Feynman diagram for the Fan-Migdal self energy, $\Sigma_\text{FM}$. The solid blue line is the electron Green's function; the wavy red line is the phonon Green's function; the black disks are the electron-phonon coupling matrix elements. The inset illustrates the Dyson equation leading to the screening of the electron-phonon matrix element; the black circle, denoted $\Gamma$, is the vertex function and the dotted black line is the bare Coulomb interaction. (b) Comparison between DFPT and GWPT calculations of the imaginary part of the Fan-Migdal self-energy, for the lowest conduction band of GaAs. The annotations indicate the values of $\text{Im}\,{\Sigma}_\text{FM}$ for thermal electrons (40~meV above the band bottom at room temperature).}
    \label{fig:1}
\end{figure*}

The scattering rate of an electron by phonons is obtained from the phonon-induced Fan-Migdal self-energy $\Sigma_{\rm FM}$, which is shown in Fig.~\ref{fig:1}(a) via its Feynman diagram representation \cite{giustino2017electron}. The imaginary part of this self-energy corresponds to the spectral broadening of the electron quasiparticle due to electron-phonon interactions, and relates to its lifetime $\tau$ via ${\rm Im}\,\Sigma_{\rm FM} = \hbar / 2\tau$, where $\hbar$ is the reduced Planck constant. The inset of Fig.~\ref{fig:1}(a) illustrates the screening of the bare electron-phonon vertex, corresponding to the matrix element of the variation of the ionic potential in the empty lattice, by the electronic polarization. Formally, this process translates into the relation $g = \epsilon^{-1} g_{\rm b}$, where $g$ and $g_{\rm b}$ are the screened and bare vertices, respectively, and $\epsilon$ is the dielectric matrix resulting from electronic Hartree screening as well as exchange and correlation effects. In standard DFT and density-functional perturbation theory (DFPT) \cite{baroni2001phonons}, the many-body dielectric matrix is effectively replaced by the DFT dielectric matrix; this approximation tends to overscreen the bare electron-phonon vertex when using semi-local functionals~\cite{giustino2014materials}. By consequence, DFT and DFPT approaches tend to underestimate the electron-phonon coupling strength as compared to experiments \cite{li2019electron, lazzeri2008impact, antonius2014many, monserrat2016correlation}. Beyond the electron-phonon vertex, DFT calculations may not describe band dispersions and band effective masses with sufficient accuracy~\cite{golze2019gw}, introducing additional errors in the evaluation of the Fan-Migdal self-energy through the Green's function $G$ shown as a straight line in Fig.~\ref{fig:1}(a).

Figure~\ref{fig:1}(b) illustrates the importance of many-body corrections for the case of GaAs. Near the conduction band minimum, the imaginary part of the phonon-induced self-energy is 3.3$\times$ larger with GW bands and GWPT matrix elements as compared to conventional DFT/DFPT; this enhancement results from an increase of the band effective mass by 89\%, and an increase of the electron-phonon coupling matrix elements by 14\%. Since the carrier mobility is directly proportional to the lifetime and inversely proportional to the mass (see Supplemental Methods), these corrections are expected to significantly impact the mobility computed from DFT/DFPT and cannot be neglected in predictive calculations.

Supplemental Figure~S1 provides a schematic summary of the calculation workflow~\cite{SM_note}. The matrix element $g^\text{DFT}_{mn\nu}(\textbf{k},\textbf{q}) = \bra{\psi_{m\textbf{k}\!+\!\textbf{q}}} \Delta_{\textbf{q}\nu} V_\text{SCF} \ket{\psi_{n\textbf{k}}}$ is the probability amplitude for an electron in the state $\ket{\psi_{n\textbf{k}}}$ with wavevector $\textbf{k}$ and band $n$ to scatter into the state $\ket{\psi_{m\textbf{k}\!+\!\textbf{q}}}$ via a phonon with wavevector $\textbf{q}$ in the branch $\nu$. $\Delta_{\textbf{q}\nu}V_\text{SCF}$ denotes the variation of the DFT self-consistent potential with respect to the atomic displacements associated to this phonon, and contains two contributions: the Hartree contribution, $\Delta_{\textbf{q}\nu}V_\text{H}$, and the exchange and correlation contribution, $\Delta_{\textbf{q}\nu}V_\text{xc}$. GWPT corrects the overscreening of the electron-phonon vertex by replacing the latter with the variation of the many-body GW self-energy, $\Delta_{\textbf{q}\nu} \Sigma_{\rm GW}$: $g^\text{GW}_{mn\nu}(\textbf{k},\textbf{q}) = g^\text{DFT}_{mn\nu}(\textbf{k},\textbf{q}) + \bra{\psi_{m\textbf{k}\!+\!\textbf{q}}} \Delta_{\textbf{q}\nu} (\Sigma_{\rm GW}-V_\text{xc}) \ket{\psi_{n\textbf{k}}} $ \cite{li2019electron,li2024electron}. For each $\textbf{q}$-point, these corrections require $3N$ calculations of $\Delta \Sigma_{\rm GW}$, where $N$ is the number of atoms in the unit cell; this step is the most computationally demanding part of the workflow (see Supplemental Methods~\cite{SM_note}).

The evaluation of the scattering lifetimes needed in the \aibte\ needs to be performed on ultra-dense grids of $(\textbf{k},\textbf{q})$ wavevectors in the Brillouin zone, of the order of 100$^3$ points \cite{ponce2021first}; however, GWPT calculations on such dense grids are prohibitive. To circumvent this difficulty, we combine GWPT with electron-phonon Wannier interpolation \cite{giustino2017electron, li2019electron} as implemented in the EPW code \cite{lee2023electron}. From the quasiparticle energies and band velocities, together with the many-body scattering rates, we solve the \aibte\ using the Jacobi iteration method \cite{ponce2018towards}.

\begin{figure*}
    \includegraphics{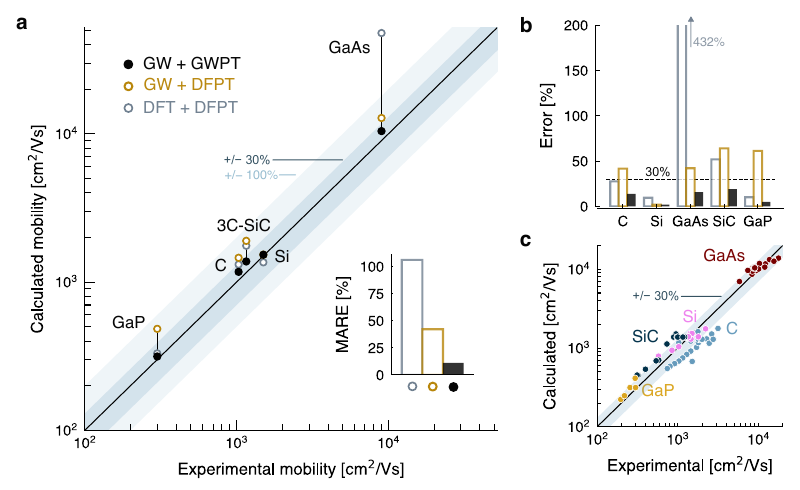}
    \caption{(a) Comparison between the Hall mobilities of GaP, 3C-SiC, Si, diamond, and GaAs calculated within DFT+DFPT (blue), GW+DFPT (orange), and GW+GWPT (black) with experimental data. All data refer to room temperature and nominally undoped or weakly doped samples (doping concentration $\leq 10^{16}$~cm$^{-3}$). The inset shows the mean absolute relative error (MARE). (b) Absolute relative error between theory and experiment, by semiconductor. (c) Comparison of GWBTE mobilities with experimental data of nominally undoped semiconductors at various temperatures and different processing and measurement conditions (including drift and Hall measurements). The color code is: orange for GaP, dark blue for 3C-SiC, pink for Si, light blue for diamond, and red for GaAs.} 
    \label{fig:2}
\end{figure*}

Figure~\ref{fig:2}(a) compares our calculated mobilities with experimental data. We focus on five prototypical semiconductors, namely Si~\cite{sze2008semiconductor}, GaAs~\cite{hicks1969high}, GaP~\cite{haynes2020crc}, diamond~\cite{pernot2006hall}, and 3C-SiC~\cite{nelson1966growth}, which are central to modern electronics and can be synthesized with exceptionally high crystal quality; this choice allows us to benchmark \textit{intrinsic} phonon-limited transport properties. For the same reason, we focus on nominally undoped and weakly doped single-crystal samples at 300\,K (measured carrier concentrations are given in Supplemental Tab.~S1). We consider three calculation levels of increasing sophistication: (i) DFT bands and DFPT electron-phonon couplings; (ii) GW bands and DFPT electron-phonon couplings; and (iii) GW bands and GWPT electron-phonon couplings; in the following, we will refer to this fully many-body approach as ``GWBTE''. In Fig.~\ref{fig:2}(a) we see how the inclusion of GW self-energy effects in \textit{both} electron bands and electron-phonon couplings delivers the most accurate results compared to experimental results (i.e., data points are closer to the 45$^\circ$ line). The inset of the same panel shows that, while standard DFT calculations yield a mean absolute relative error (MARE) of 106\%, and GW bands reduce this error to 42\%, the full-blown calculation including \bb{GW self-energy effects} on both bands and matrix elements reaches an impressive 11\% MARE [cf.\ Supplemental Fig.~S2 for temperature dependence, Supplemental Fig.~S3 for convergence tests, and Supplemental Tab.~S2 for the data points in Fig.~\ref{fig:2}(a)]. \vv{These errors should not be interpreted as indicative of the intrinsic statistical accuracy of the underlying methods, which requires systematic evaluation over a broader and more diverse dataset.}

Figure~\ref{fig:2}(b) presents a quantitative analysis of these corrections for each material. The surprising result is that the widely adopted approach of combining GW-corrected band structures with DFPT-level electron-phonon couplings does not necessarily yield better agreement with experiment. In fact, for diamond, SiC, and GaP, this hybrid scheme worsens the predictions relative to using DFT throughout. In these cases, the GW correction slightly reduces the band effective mass, which would tend to increase the mobility; however, the many-body corrections to the electron-phonon interaction are much more pronounced, ultimately lowering the mobility. \bb{Supplemental Tab.~S3 shows a cross-validation of the error statistics by systematically omitting one material and recomputing the average error. This analysis shows that the GW+GWPT error always remains less than 15\% and at least two times smaller than DFT+DFPT and GW+DFPT errors.} These findings highlight the need for including \bb{GW self-energy effects} in \textit{both} the band structure and the electron-phonon coupling to achieve predictive accuracy. Relying on GW bands alone, as done in most transport calculations to date, can be misleading.

To further validate our analysis, in Fig.~\ref{fig:2}(c) we report our GWBTE calculations against 66 experimental measurements spanning various growth methods, temperatures, and transport measurements. In particular, we compile data for Si \cite{jacoboni1977review, norton1973impurity, madelung2004semiconductors, ludwig1956drift, sze2008semiconductor}, GaAs~\cite{hicks1969high, rode1971electron, stanley19914, lin1982vapour, madelung2004semiconductors}, GaP~\cite{miyauchi1967electrical, madelung2004semiconductors}, 3C-SiC~\cite{nelson1966growth, shinohara1988growth, mnatsakanov2001semiempirical, bhatnagar1993ieee}, and diamond~\cite{vavilov1976semiconducting, madelung2004semiconductors, haynes2020crc, gabrysch2011electron, jansen2013temperature, nesladek2008charge, nava1980electron, redfield1954electronic}. We restrict our attention to temperatures above 250\,K, where phonon scattering dominates over impurity scattering~\cite{leveillee2023ab}. Full details of the measurement conditions, along with the experimental and calculated mobilities, are provided in Supplemental Tab.~S4. As shown in Fig.~\ref{fig:2}(c), our GWBTE calculations closely track the experimental values, where each theoretical value corresponds to a different experimental temperature. The MARE over this diverse dataset is 24\%; this error is slightly higher than the 11\% reported above, but we point out that the experimental data exhibit significant scatter, therefore this metric reflects both theoretical and \textit{experimental} uncertainties~\cite{diamond_note}.

\begin{figure*}[htp!]
    \includegraphics{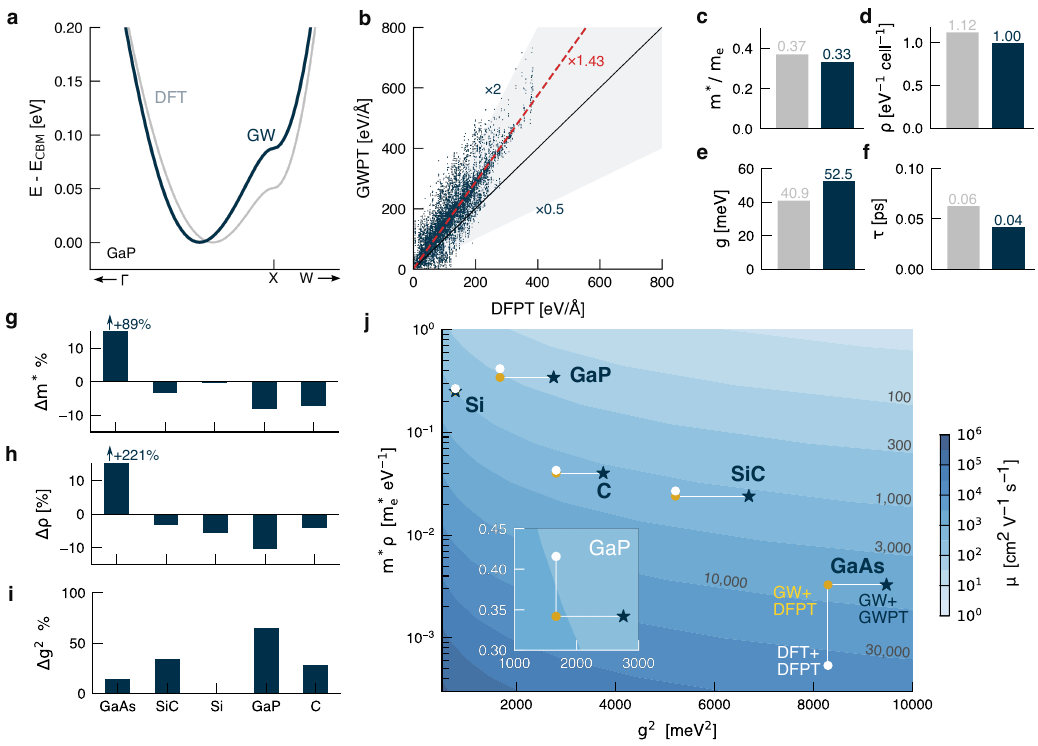}
    \caption{(a) Comparison between DFT (gray) and GW (blue) band structure of GaP.
    (b) Comparison between \bb{the} electron-phonon matrix elements \bb{calculated at the DFT level (DFPT) and the GW level (GWPT)} in GaP (dots) for states within 1.0 eV of the conduction-band bottom. The red line indicates the average ratio of 1.43; the black line indicates a 1:1 ratio, for comparison; the shading indicates the region where GWPT matrix elements are within a factor $2\times$ of the corresponding DFPT matrix elements. For convenience, in this panel we show the magnitude of the matrix elements in the basis of atomic displacements, $|\bra{\psi_{m\textbf{k}+\textbf{q}}} \Delta_{\kappa\alpha,\textbf{q}} V \ket{\psi_{n\textbf{k}}}|$, where $\kappa$ is the atom label and $\alpha$ is the Cartesian direction~\cite{giustino2017electron}. These matrix elements are evaluated on coarse grids across the Brillouin zone.
    (c)-(f) Comparison between DFT (gray) and GW (blue) calculations of the effective parameters $m^*$, $g$, $\rho$, and $\tau$ defined in the main text, for GaP. The average increase of electron-phonon coupling for states across the Brillouin zone (43\% in (b)) is generally not equivalent to the increase of $g$ relevant for band-edge states that participate in transport (28\% in (e)). (g)-(i)Many-body corrections to the effective parameters $m^*$, $\rho$, $g^2$ across all materials considered.
    (j) Two-dimensional map of \bb{GW self-energy effects} on the electron mobility with the vertical axis in logarithmic scale. The inset provides a zoomed-in view of the map for GaP with the vertical axis in linear scale. Contour lines are curves of constant mobility; line segments connect DFT+DFPT (white symbols), GW+DFPT (orange symbols), and GW+GWPT (black symbols) results.
    }
    \label{fig:3}
\end{figure*}

In Fig.~\ref{fig:3}, we analyze how GWBTE improves mobility calculations by disentangling the effects of many-body band structure renormalization and many-body corrections to the DFT electron-phonon coupling. Panel (a) shows how GW corrections make the bands of GaP more dispersive; this effect is expected to alter the electron velocities appearing in the \aibte, as well as the phase space for scattering in the relaxation times, cf.\ Eq.~(S7) of the Supplemental Methods. Panel (b) shows how the electron-phonon coupling matrix elements are significantly enhanced by \bb{GW self-energy effects}, with an enhancement of 43\% on average. Interestingly, these corrections have a nontrivial dependence on wavevector and energy, as it is seen from the broad distribution of the data points around the average. This complex trend cannot be captured by a simple rescaling of the DFPT matrix elements \cite{ponce2018towards}: as a visual guide, the shaded area in Fig.~\ref{fig:3}(b) shows how the corrections are widely distributed within a factor of $2\times$ above and below the DFT values. This variability highlights the importance of explicitly computing electron-phonon couplings at the GW level.

In order to systematically quantify how these corrections impact transport properties, we write the mobility $\mu$ as follows:
\begin{equation}\label{eq:1}
\mu = \frac{e\, \tau}{m^*}, \qquad \frac{1}{\tau} = \frac{2\pi}{\hbar} g^2 \rho~,
\end{equation}
where $e$ is the electron charge and $m^*$ is the conductivity effective mass. $\tau$ is the average relaxation time, and is expressed via the average electron-phonon coupling matrix element $g$ and the average density of states $\rho$. Equation~\eqref{eq:1} is intentionally cast in the form of the elementary Drude model, but it is an \textit{exact} rewriting of the full \textit{ab initio} GWBTE mobility within the self-energy relaxation time approximation. Explicit expressions for the effective quantities $m^*$, $g$, and $\rho$, derived from first-principles band structures, phonon dispersions, and electron-phonon matrix elements, are provided in the Supplemental Methods. It is important to note that Eq.~\eqref{eq:1} is valid only within the relaxation time approximation. To verify the validity of this assumption for the present analysis, we benchmark against full GWBTE calculations in Supplemental Tab.~S5. \bb{All mobilities reported up to this point have been obtained from fully iterative \aibte\ solutions. In Fig.~\ref{fig:3}, we use the self-energy relaxation time approximation solely to disentangle the effects of GW self-energy corrections on the mobility.}

Figure~\ref{fig:3}(c)-(f) illustrates, for the case of GaP, how each parameter $m^*$, $\rho$, and $g$ changes when going from DFT+DFPT to GW+GWPT. The dominant effect is the increase of the electron–phonon matrix element ($28\%$), amplified by the quadratic dependence on $g$ in Eq.~\eqref{eq:1}, while the reduction in effective mass ($12$\%) plays a secondary role. Similar analyses are shown for 3C-SiC, Si, diamond, and GaAs in Supplemental Fig.~S4. Figures~\ref{fig:3}(g)-(i) summarize these trends across all materials considered: in GaAs, mobility corrections are driven mainly by improvements in the band structure, whereas in SiC, GaP, and diamond they are primarily driven by the enhancement of the electron-phonon coupling. For silicon, \bb{GW self-energy effects} beyond DFT are negligible.

To unify these observations, we construct in Fig.~\ref{fig:3}(j) a mobility map using $g^2$ and $\rho m^*$ as descriptors. The horizontal axis quantifies the electron-phonon coupling strength, while the vertical axis serves as a proxy for the effective mass (since $\rho$ depends on $m^*$). Contour lines indicate curves of constant mobility, and line segments connect DFT+DFPT and GW+GWPT results. This representation shows how many-body corrections to electron-phonon coupling, in addition to GW band corrections, are significant for nearly all materials (GaAs, SiC, GaP, diamond) while the effects are marginal for silicon; this latter observation explains why earlier DFT-level calculations for silicon already matched experiments closely~\cite{ponce2018towards}.

Taken together, our findings demonstrate that predictive calculations of phonon-limited carrier mobility require many-body corrections to both the electronic band structure \textit{and} the electron-phonon coupling. By combining GW perturbation theory with \textit{ab initio} Boltzmann transport simulations, we achieve a fourfold improvement in mobility predictions compared to the prior state of the art. These results resolve a long standing limitation of DFT based transport methods and establish a general framework for reliable carrier mobility predictions. \bb{They also motivate a systematic reexamination of mobility calculations across a wider range of materials to more accurately quantify the errors at each level of theory, by incorporating GW self-energy corrections into both the band structure and the electron–phonon vertex.} We anticipate that extending this approach to emerging quantum materials based on topological semimetals, correlated oxides, and  Moir\'e materials will deepen our understanding of charge transport across diverse classes of solids.

\vspace{10pt}
This work was primarily supported by the Computational Materials Science program of the U.S. Department of Energy, Office of Science, Basic Energy Sciences, through award no. DE-SC0020129 (project design, calculations, data analysis, EPW development); and by the U.S. National Science Foundation through CSSI award no. OAC-2513830 (interface between EPW and BerkeleyGW). Support is also acknowledged by the Center for Computational Study of Excited-State Phenomena in Energy Materials (C2SEPEM) at the Lawrence Berkeley National Laboratory (LBNL), which is funded by the U.S. Department of Energy (DOE), Office of Science, Basic Energy Sciences, Materials Sciences and Engineering Division under Contract No. DEAC02-05CH11231, as part of the Computational Materials Sciences Program, which provided advanced codes and computation of electron-phonon couplings based on GWPT. Z.L. also acknowledged the support from the U.S. National Science Foundation through CAREER Award under Grant No. DMR-2440763 (development of GWPT new features). Computational resources were provided by the National Energy Research Scientific Computing Center (a DOE Office of Science User Facility supported under Contract No.~DE-AC02-05CH11231), the Argonne Leadership Computing Facility (a DOE Office of Science User Facility supported under Contract DE-AC02-06CH11357), and the Texas Advanced Computing Center (TACC) at The University of Texas at Austin. 

The data supporting the findings of this Letter are openly available~\cite{pant2025MatCloud}. The codes used in this work, namely \texttt{EPW}, \texttt{Quantum ESPRESSO}, \texttt{ABINIT}, \texttt{Wannier90}, and \texttt{BerkeleyGW} are all open-source software and are freely available from their respective websites. The software patch for these codes are available on GitLab~\cite{pant2025gitlab}. \nocite{giannozzi2017advanced, giannozzi2020quantum, hamann2013optimized, van2018pseudodojo, perdew1996generalized, ceperley1980ground, perdew1981self, Pizzi2020, autowann2025, verdi2015frohlich, brunin2020electron, deslippe2012berkeleygw, ponce2016epw, STILLMAN19701199}.

\clearpage
\newpage


%

\end{document}


\title{Supplemental materials\\[5pt]
First-principles predictions of carrier mobility with record accuracy\\ using GW perturbation theory}
\author{Nick Pant}
\affiliation{\oden} 
\affiliation{\utphysics} 

\author{Sabyasachi Tiwari}
\affiliation{\oden} 
\affiliation{\utphysics} 

\author{Steven G. Louie}
\affiliation{\ucphysics}
\affiliation{\lbl} 

\author{Zhenglu Li}
\affiliation{\usc} 

\author{Feliciano Giustino}
\affiliation{\oden} 
\affiliation{\utphysics} 

\maketitle
\onecolumngrid

\setlength{\parskip}{3pt}

\begin{tabular}{l@{\hskip 10pt}l}
\textbf{Contents} & \\[4pt]
Supplemental Methods 
 & Ground state calculations\\[2pt]
 & Wannier-Fourier interpolation\\[2pt]
 & Quasiparticle calculations\\[2pt]
 & GW perturbation theory calculations\\[2pt]
 & Drude-like factorization of the \textit{ab initio} mobility\\[2pt]
Supplemental Table~S1   & Reported carrier concentrations in experiments\\[2pt]
Supplemental Table~S2   & Data points in Fig. 2(a)\\[2pt]
\bb{Supplemental Table~S3}   & \bb{Cross-validation for mobility error statistics}\\[2pt]
Supplemental Table~S4   & Data points in Fig. 2(c) and provenance\\[2pt]
Supplemental Table~S5   & Comparison between BTE and SERTA mobilities\\[2pt]
\bb{Supplemental Table~S6}   & \bb{Full-frequency calculation of the GW electron--phonon interaction.}\\[2pt]
Supplemental Table~S7   & Comparison between theoretical and experimental band gaps\\[2pt]
\bb{Supplemental Table~S8}   & \bb{Convergence of GW calculations in diamond}\\[2pt]
\bb{Supplemental Table~S9}   & \bb{Convergence of GWPT calculations in diamond}\\[2pt]
\bb{Supplemental Table~S10}   & \bb{Computational cost of GWPT calculations}\\[2pt]
Supplemental Figure~S1  & Schematic workflow of GWBTE calculation\\[2pt]
Supplemental Figure~S2  & Temperature dependence of mobilities\\[2pt]
Supplemental Figure~S3  & Convergence tests for mobility\\[2pt]
Supplemental Figure~S4  & Disentangling mobility descriptors\\[2pt]
Supplemental Figure~S5  & Decay of electron-phonon matrix elements in the Wannier representation\\
\bb{Supplemental Figure~S6} & \bb{Full-frequency calculation of GW energies.}\\[2pt]
\end{tabular}

\newpage
\textbf{Supplemental methods}

\smallskip

\textbf{Ground-state calculations.} We performed ground-state DFT and DFPT calculations using \texttt{Quantum ESPRESSO}~\cite{giannozzi2017advanced, giannozzi2020quantum}. We relaxed the lattice constants and the internal atomic coordinates by enforcing all forces to be below $10^{-3}$~Ry/bohr and pressures to be below 0.5~kbar. We computed the ground-state charge density with Monkhort-Pack BZ sampling grids of $12^3$ \textbf{k} points for diamond, Si, and 3C-SiC, and $14^3$ points for GaP and GaAs. We used optimized norm-conserving Vanderbilt pseudopotentials~\cite{hamann2013optimized} obtained from the \texttt{Pseudo Dojo} library~\cite{van2018pseudodojo}, employing plane-wave kinetic-energy cutoffs of 80 Ry for Si, 3C-SiC, and GaP, 100 Ry for diamond, and 130 Ry for GaAs. For the exchange-correlation functional, we used the generalized-gradient approximation of Perdew, Burke and \bb{Ernzerhof} for Si, 3C-SiC, and diamond~\cite{perdew1996generalized}, and the Perdew-Zunger parameterized local-density approximation for for GaAs and GaP~\cite{ceperley1980ground, perdew1981self}; with these choices, GW-corrected band gaps are in close agreement with experiments. The computed charge densities were then used to compute lattice dynamical quantities such as phonon frequencies and electron-phonon matrix elements using DFPT~\cite{baroni2001phonons}.

\smallskip

\textbf{Wannier-Fourier interpolation.}
Mobility calculations require the evaluation of electron energies, phonon frequencies, and electron-phonon matrix elements on dense BZ sampling grids. To this end, we used Wannier-Fourier interpolation of bands as implemented in the \texttt{Wannier90} code~\cite{Pizzi2020}, and of electron-phonon matrix elements and band velocities as implemented in the \texttt{EPW} code~\cite{lee2023electron}. We employed electronic coarse grids of $4^3$ \textbf{k} points for GaAs, and $8^3$ \textbf{k}-points for Si, 3C-SiC, GaP, and diamond. Likewise, we used phonon coarse grids of $4^3$ \textbf{q}-points for GaAs, Si, 3C-SiC, and GaP, and $8^3$ \textbf{q}-points for diamond. We used a meta-optimization scheme to generate maximally-localized Wannier functions~\cite{autowann2025}. For long-range electron-phonon couplings, we employed dipole and quadrupole corrections at the DFPT level~\cite{verdi2015frohlich, brunin2020electron}.
Dynamical quadrupoles were \bb{computed} using the \texttt{ABINIT} code~\cite{gonze2020abinit}.
Supplemental Fig.~S5 shows the decay of the short-range part of the electron-phonon matrix elements in the Wannier representation. \bb{This procedure yields accurate mobility predictions for both polar and nonpolar semiconductors compared to experiments, and we expect that a recently developed approach to calculate the long-range GWPT Fr\"{o}hlich interaction would further improve the accuracy of interpolation for polar materials~\cite{zhu2025many}. We interpolated the GW energies using the unitary rotation matrices obtained from DFT wave functions, and computed the velocity matrix elements through the derivatives of the Hamiltonian in the Wannier representation, according to Eq.~(31) of ref.~\cite{wang2006ab}.} 

\smallskip

\textbf{Quasiparticle corrections.}
We applied many-body quasiparticle corrections to the electron energies and electron-phonon matrix elements within the $G_0W_0$ approximation using \texttt{BerkeleyGW}~\cite{deslippe2012berkeleygw}, within the Hybertsen-Louie plasmon-pole approximation~\cite{hybertsen1986electron}. \bb{We compare this approximation to a full-frequency calculation in Supplemental Fig.~S6 for the GW energies and in Supplemental Tab.~S6 for the electron--phonon interaction.} To converge the dielectric function, we used a kinetic energy cutoff of 35~Ry for all materials. For the summation over empty bands, we includeds empty states with energies up to 26~Ry for diamond, 15~Ry for GaAs, 13~Ry for GaP, 10~Ry for Si, and 20~Ry for 3C-SiC, all referenced to the respective valence band maxima. We compare the calculated band gaps to experimental band gaps in Supplemental~Tab.~S7. \bb{In evaluating the GW self-energy, we approximated the vertex function with a delta function, $\Sigma = iGW\Gamma \approx iGW$~\cite{hybertsen1986electron}. Further implementation details including the treatment of the Coulomb singularity can be found in Ref.~\citenum{deslippe2012berkeleygw}. For a detailed review on the limitations and successes of the GW approximation, including dependence on the starting-point calculation, we refer the reader to Refs.~\citenum{reining2018gw} and \citenum{golze2019gw}. We show convergence tests for the conductivity effective mass and the direct band gap in Supplemental Tab. S8. }

\smallskip

\textbf{GW perturbation theory calculations.} For GWPT calculations, we employed the method of Refs.~\citenum{li2019electron,li2024electron}. In this method, the first-order change of the GW self-energy due to the displacement of the sublattice $\kappa$ in the direction $\alpha$ is expressed in terms of the variation of the Green's function as:
\begin{equation}
    \Delta_{\textbf{q}\kappa\alpha} \Sigma = i \int \frac{d\omega}{2\pi} e^{-i\delta\omega} \Delta_{\textbf{q}\kappa\alpha} G(\textbf{r},\textbf{r}'; \varepsilon - \omega) W(\textbf{r},\textbf{r}';\omega).
\end{equation}
In turn, the variation of the Green's function is expressed via the linear response of Kohn-Sham states:
These states are used to construct the linear response of the Green's function,
\begin{equation}
    \Delta_{\textbf{q}\kappa\alpha} G(\textbf{r},\textbf{r}';\varepsilon) = \sum_{n\textbf{k}} \frac{\Delta_{\textbf{q}\kappa\alpha}\psi_{n\textbf{k}}(\textbf{r}) \psi_{n\textbf{k}}^*(\textbf{r}')  + \psi_{n\textbf{k}}(\textbf{r}) [\Delta_{\textbf{q}\kappa\alpha} \psi_{n\textbf{k}}(\textbf{r}')]^*}{\varepsilon - \varepsilon_{n\textbf{k}} - i\delta_{n\textbf{k}}},
\end{equation}
where $\delta_{n\textbf{k}} = 0^\pm$ for empty/occupied states, respectively. 
These calculations are demanding because they require the linear response $\Delta_{\textbf{q}\kappa\alpha} \psi_{n\textbf{k}}$ for all phonon perturbations, each needing a summation over hundreds of empty Kohn-Sham states. The variations of the Kohn-Sham states are obtained from the \texttt{ABINIT} code~\cite{gonze2020abinit}. \bb{In constructing the linear response of the Green's function, we neglect the variation of the eigenvalue because the first-order eigenvalue correction is exactly zero for finite-\textbf{q} phonons due to the conservation of crystal momentum, \textit{i.e.}, $\bra{n\textbf{k}} \Delta_{\textbf{q}\nu} V \ket{n\textbf{k}} = 0$, as discussed in the Supplemental Materials of Ref.~\citenum{li2019electron}. For phonon perturbations at $\textbf{q}=0$, from a Brillouin zone (BZ) sampling point of view, the $\textbf{q}=0$ point represents a small finite-volume region in the BZ, where the exact q=0 occupies infinitesimal phase space in 3D, and for the region represented by $\textbf{q}\neq0$, the first-order changes of eigenvalues are still rigorously zero. Therefore, we did not include such effects. Moreover, we also neglect the variation of the screened Coulomb interaction which has been widely tested to be valid, known as the constant-screening approximation, as demonstrated in Ref.~\citenum{faber2015exploring} This method applies generally to materials that are weakly to moderately correlated and also to both isotropic and anisotropic materials. Strong correlations require higher-order diagrams beyond the first-order GW approximation.} 

In this work, GWPT corrections are applied to the lowest conduction band of C, GaAs, and SiC and the lowest two conduction bands of GaP and Si. \bb{We show convergence tests for the GWPT electron--phonon matrix elements of diamond in Supplemental Tab.~S9. GWPT calculations scale with the number of atoms $N$, the number of \textbf{k}-points $N_\textbf{k}$, and number of \textbf{q}-points $N_\textbf{q}$ as $N_\textbf{k} N_\textbf{q} N^5\log N $. The calculation of the plane-wave matrix elements scales with the number of plane waves $N_G$ as $N_G^2 \log N_G$, which is repeated per phonon mode, yielding another factor of $N_\text{modes}$. If this is performed for all diagonal and off-diagonal matrix elements, then we obtain an additional factor of $N_\text{band}^2$. Since the calculation is performed for a grid of ($\textbf{k}$, $\textbf{q}$) points, we obtain a factor of $N_\textbf{k}N_\textbf{q}$. Recognizing that $N_G$, $N_\text{bands}$, and $N_\text{modes}$ all scale linearly with the number of atoms $N$, we obtain the scaling form shown above. The cost of the GWPT calculations for the materials shown here are given in Supplemental Tab.~S10. We anticipate that GWPT calculations can be extended to larger classes of materials in the future by combining GPU acceleration~\cite{zhang2025advancing} with machine-learning approaches.}

\smallskip

\textbf{Mobility calculations.} 
To compute the electron mobility, we solved the \textit{ab initio} linearized Boltzmann transport equation (\textit{ai}BTE) using the Jacobi iteration method, as implemented in the \texttt{EPW} code~\cite{ponce2016epw, lee2023electron}:
\begin{eqnarray}
    &&\Big[1 - \frac{e}{\hbar} \tau_{n\textbf{k}} (\textbf{v}_{n\textbf{k}} \cross (b)) \cdot \nabla_\textbf{k} \Big] 
    \partial_{E_\beta} f_{n\textbf{k}} = e {v}_{n\textbf{k}\beta} \frac{\partial f^0_{n\textbf{k}}}{\partial \varepsilon_{n\textbf{k}}} \tau_{n\textbf{k}}
    + \frac{2\pi\tau_{n\textbf{k}}}{\hbar} \sum_{m\nu} \int \frac{d\textbf{q}}{\Omega_\text{BZ}} \abs{g_{mn\nu}(\textbf{k},\textbf{q})}^2 \nonumber \\
    && \hspace{30pt}\times \big[(n_{\textbf{q}\nu} + 1 - f^0_{n\textbf{k}})\delta(\varepsilon_{n\textbf{k}} - \varepsilon_{m\textbf{k}+\textbf{q}} + \hbar\omega_{\textbf{q}\nu}) +(n_{\textbf{q}\nu} + f_{n\textbf{k}}^0)\delta(\varepsilon_{n\textbf{k}} - \varepsilon_{m\textbf{k}+\textbf{q}} - \hbar\omega_{\textbf{q}\nu})\big] \partial_{E_\beta} f_{m\textbf{k}+\textbf{q}}. \label{eq:aibte}
\end{eqnarray}
The quantities appearing in this equation, from left to right, have the following meaning. $e$ and $\hbar$ are the electron charge and Planck constant, respectively; $\tau_{n\textbf{k}}$ is the electron relaxation time, as defined in Eq.~\eqref{eq:serta_tau} below; $\textbf{v}_{nk}$ is the band velocity; $(b)$ denotes a finite magnetic field, taken in this work to be $10^{-10}$~T to capture the linear response of the system.  $\partial_{E_\beta} f_{n\textbf{k}}$ is the linear response of the occupation function to an electric field $(e)$ along the Cartesian direction $\beta$. $f^0_{n\textbf{k}}$ is the equilibrium Fermi-Dirac occupation. $\Omega_\text{BZ}$ is the volume of the Brillouin zone, $g_{mn\nu}(\textbf{k},\textbf{q})$ are the electron-phonon matrix elements, $\varepsilon_{n\textbf{k}}$ are electron energies, and 
$\omega_{\textbf{q}\nu}$ are phonon frequencies. $n_{\textbf{q}\nu}$ is the Bose-Einstein occupation factor. \bb{In our computational workflow, we use the electron energies and phonon frequencies calculated with \texttt{Quantum ESPRESSO}, and the electron--phonon matrix elements computed with \texttt{ABINIT}, while ensuring that identical structures and settings are used for all calculations.} From the solution $\partial_{E_\beta} f_{n\textbf{k}}$ of this equation, we obtain the drift and Hall mobilities by evaluating the linear response of the electric current, as detailed in Ref.~\citenum{lee2023electron}.

In the present work, we set the carrier concentration to a very low value, $n_{\rm c}=10^{13}$~cm$^{-3}$, to capture the near-intrinsic limit. At this concentration, the mobility is essentially independent of the concentration. We focus on the electron mobility because hole mobilities require spin-orbit coupling~\cite{ponce2021first}, which is not yet implemented in the GWPT framework. \bb{Our calculations do not include higher order corrections to the atomic structure, electron bands, and phonon frequencies from finite-temperature renormalization. To account for temperature dependence, we computed the mobility at temperatures ranging from 200 K to 500 K in 50 K increments and used spline interpolation to obtain mobilities at intermediate temperatures. The interpolation order was found to have no appreciable effect on the resulting values.}

\smallskip

\textbf{Drude-like factorization of the \textit{ab initio} mobility}. Here, we outline the procedure for factorizing the \textit{ab initio} mobility into a Drude-like form. We work with the mobility in the self-energy relaxation-time approximation (SERTA) to keep the influence of the BTE iterations separate from the impact of $GW$ corrections on the overall mobility. The SERTA mobility is given by:
\begin{equation}
    \mu_{\alpha\beta} = -\frac{e}{\Omega_\text{uc}n_{\rm c}} \sum_n \int \frac{d\textbf{k}}{\Omega_\text{BZ}} \frac{\partial f^0_{n\textbf{k}}}{\partial \varepsilon_{n\textbf{k}}} v_{n\textbf{k}\alpha} v_{n\textbf{k}\beta} \tau_{n\textbf{k}}, \label{eq:serta}
\end{equation}
where $\Omega_\text{uc}$ is the volume of the crystal unit cell, and the relaxation time $\tau_{n\textbf{k}}$ is given by the Fermi golden rule:
\begin{equation}
    \tau_{n\textbf{k}}^{-1} = \frac{2\pi}{\hbar} \sum_{m\nu} \int \frac{d\textbf{q}}{\Omega_\text{BZ}} \abs{g_{mn\nu}(\textbf{k},\textbf{q})}^2  \Big[ (n_{\textbf{q}\nu} + 1 - f^0_{m\textbf{k}+\textbf{q}})\delta(\varepsilon_{n\textbf{k}} - \varepsilon_{m\textbf{k}+\textbf{q}} - \hbar\omega_{\textbf{q}\nu}) 
    + (n_{\textbf{q}\nu} + f^0_{m\textbf{k}+\textbf{q}})\delta(\varepsilon_{n\textbf{k}} - \varepsilon_{m\textbf{k}+\textbf{q}} + \hbar\omega_{\textbf{q}\nu}) \Big]. \label{eq:serta_tau}
\end{equation}
To connect \textit{ab initio} calculations with the Drude model, where the mobility is expressed as $\mu = e \tau / m^*$, we use the conductivity effective mass, which is the band effective mass averaged over the states that contribute to the current.~\cite{madsen2006boltztrap} Formally, this mass can be obtained from Eq.~\eqref{eq:serta} by replacing $e \tau_{n\textbf{k}}$ with unity:
\begin{equation}
    (m^*)^{-1}_{\alpha\beta} = -\frac{1}{\Omega_\text{uc}n_{\rm c}} \sum_n\int \frac{d\textbf{k}}{\Omega_\text{BZ}} \frac{\partial f^0_{n\textbf{k}}}{\partial \varepsilon_{n\textbf{k}}} v_{n\textbf{k}\alpha} v_{n\textbf{k}\beta}. \label{eq:mass}
\end{equation}
 Using Eqs.~\eqref{eq:serta} and \eqref{eq:mass}, and taking the isotropic average of both tensors since all compounds considered in this work have cubic symmetry, we can define the average relaxation time by inverting the following relation: 
\begin{equation}
    \mu = \frac{e \tau}{m^*}, \label{eq:crta}
\end{equation}
which is equivalent to evaluating the average: 
\begin{equation}
    \tau = \frac{\sum_{n\alpha} \int \!d\textbf{k}\, (\partial f^0_{n\textbf{k}}/\partial \varepsilon_{n\textbf{k}}) v_{n\textbf{k}\alpha}^2 \tau_{n\textbf{k}}}
    {\sum_{n\alpha}\int\! d\textbf{k}\, (\partial f^0_{n\textbf{k}}/\partial \varepsilon_{n\textbf{k}})
    v_{n\textbf{k}\alpha}^2 }. \label{eq:crta2}
\end{equation}
With these definitions, $\tau$ and $m^*$ allow us to express the \textit{ab initio} SERTA mobility in the same form as in the Drude model.

To further disentangle the contributions of electron-phonon couplings and density of states to the relaxation time $\tau$, we define a state-resolved scattering density of states $\rho_{n\textbf{k}}$ by replacing $(2\pi/\hbar)\abs{g_{mn\nu}(\textbf{k},\textbf{q})}^2$ in Eq.~\eqref{eq:serta_tau} with unity:
\begin{equation}
    \rho_{n\textbf{k}} = \sum_{m\nu} \int \frac{d\textbf{q}}{\Omega_\text{BZ}} \Big[ (n_{\textbf{q}\nu} + 1 - f^0_{m\textbf{k}+\textbf{q}})\delta(\varepsilon_{n\textbf{k}} - \varepsilon_{m\textbf{k}+\textbf{q}} - \hbar\omega_{\textbf{q}\nu}) 
     + (n_{\textbf{q}\nu} + f^0_{m\textbf{k}+\textbf{q}})\delta(\varepsilon_{n\textbf{k}} - \varepsilon_{m\textbf{k}+\textbf{q}} + \hbar\omega_{\textbf{q}\nu}) \Big]. \label{eq:rho}
\end{equation}
This state-resolved scattering density of states allows us to extract an average electron-phonon coupling strength $g$ by rewriting Eq.~\eqref{eq:crta2} as:
\begin{equation}
    \tau = \frac{1}{g^2}\frac{\hbar}{2\pi}\frac{\sum_{n\alpha} \int \!d\textbf{k}\, (\partial f^0_{n\textbf{k}}/\partial \varepsilon_{n\textbf{k}}) v_{n\textbf{k}\alpha}^2 \rho^{-1}_{n\textbf{k}} }
    {\sum_{n\alpha}\int\! d\textbf{k}\, (\partial f^0_{n\textbf{k}}/\partial \varepsilon_{n\textbf{k}})
    v_{n\textbf{k}\alpha}^2 }. \label{eq:tau}
\end{equation}
Equivalently, we can obtain an explicit expression for the average coupling strength $g$ by inverting this relation and using Eq.~\eqref{eq:crta2}:
\begin{equation}
    g^2  = 
    \frac{\hbar}{2\pi}\frac{\sum_{n\alpha} \int \!d\textbf{k}\, (\partial f^0_{n\textbf{k}}/\partial \varepsilon_{n\textbf{k}}) v_{n\textbf{k}\alpha}^2 \rho^{-1}_{n\textbf{k}} }
    {\sum_{n\alpha} \int \!d\textbf{k}\, (\partial f^0_{n\textbf{k}}/\partial \varepsilon_{n\textbf{k}}) v_{n\textbf{k}\alpha}^2 \tau_{n\textbf{k}}}. \label{eq:g2}
\end{equation}
If we also define the average scattering density of states by using the rightmost ratio in Eq.~\eqref{eq:tau}:
\begin{equation}
    \rho^{-1} = \frac{\sum_{n\alpha} \int \!d\textbf{k}\, (\partial f^0_{n\textbf{k}}/\partial \varepsilon_{n\textbf{k}}) v_{n\textbf{k}\alpha}^2 \rho^{-1}_{n\textbf{k}} }
    {\sum_{n\alpha}\int\! d\textbf{k}\, (\partial f^0_{n\textbf{k}}/\partial \varepsilon_{n\textbf{k}})
    v_{n\textbf{k}\alpha}^2 }, \label{eq:rho}
\end{equation}
we can combine Eqs.~\eqref{eq:mass} and \eqref{eq:tau}-\eqref{eq:rho} to rewrite the SERTA mobility in Eq.~\eqref{eq:serta} as follows, \textit{without approximations}: 
\begin{equation}
    \mu = \frac{e}{m^*} \left(\frac{2\pi}{\hbar} g^2 \rho \right)^{\!\!-1}~. \label{eq:factorized}
\end{equation}
We note that $\rho$ and $m^*$ are uniquely determined by the band structure, the temperature, and the chemical potential, while $g$ additionally depends on the electron-phonon coupling matrix elements.

\clearpage

\vspace*{\fill}

\begin{table}
\caption{\textbf{Carrier concentrations in experiments.} The experimental measurements of mobility reported in Fig.~2(a) of the main text correspond to nominally undoped or weakly-doped samples. Here, we indicate the measured carrier concentrations reported in the manuscripts from which the mobility measurements were taken.}
\vspace{10pt}
\centering
\begin{tabular}{@{\extracolsep{\fill}} l @{\hspace{0.5cm}} r @{\hspace{0.5cm}} r}
\hline\\[-9.5pt] \hline\\[-5pt]
Material & Carrier Concentration & Reference\\
 &(cm$^{-3}$) & \\[2pt]
&  & \\[-8pt]
\hline\\[-5pt]
GaP      & N/A &\cite{haynes2020crc} \\
3C-SiC   & $10^{16}$ &\cite{nelson1966growth} \\
Si       & $\leq 10^{12}$ &\cite{sze2008semiconductor} \\
Diamond  & $7 \times 10^{16}$ &\cite{pernot2006hall} \\
GaAs     & $2.5 \times 10^{13}$ &\cite{hicks1969high} \\[3pt]
\hline\\[-9.5pt] \hline\\[-7pt]
\end{tabular}
\end{table}

\vspace*{\fill}

\clearpage
\newpage

\begin{table}
\caption{\textbf{Calculated and experimental mobilities reported in Fig.~2(a).} Drift and Hall mobilities presented in Fig.~2(a) and related data. Theoretical values are our GWBTE calculations, extrapolated to the limit of dense Brillouin zone sampling. All data refer to room temperature. The value marked by the dagger ($\dagger$) includes a contact-size correction of +30\% determined in the experimental paper. Bold font marks the values reported in Fig.~2(a) for both theory and experiments; the experimental data in bold represent the most accurate values in the literature.}
\vspace{10pt}
\centering
\begin{tabular}{@{\extracolsep{\fill}} l r r r r}
\hline\\[-9.5pt] \hline\\[-5pt]
& \multicolumn{2}{c}{Drift Mobility} & \multicolumn{2}{c}{Hall Mobility} \\
\cmidrule(rr){2-3} \cmidrule(rr){4-5}
Material & Theory & Experiment & Theory & Experiment \\
 & (cm$^2$/Vs) & (cm$^2$/Vs) & (cm$^2$/Vs) & (cm$^2$/Vs) \\[2pt]
\hline\\[-5pt]
GaP & 271 & 
& \textbf{315} & 
\begin{tabular}[c]{@{}r@{}}256~\cite{miyauchi1967electrical}\\ \textbf{300}~\cite{haynes2020crc}\end{tabular} \\[20pt]
3C-SiC & 1265 & 
& \textbf{1377} & 
\begin{tabular}[c]{@{}r@{}}763~\cite{shinohara1988growth}\\ 890~\cite{nelson1966growth}\\ 980~\cite{nelson1966growth}\\ \textbf{1157}$^\dagger$~\cite{nelson1966growth}\end{tabular} \\[20pt]

Si & 1390 & 
\begin{tabular}[c]{@{}r@{}}1350~\cite{ludwig1956drift}\\ 1350~\cite{jacoboni1977review}\end{tabular} 
& \textbf{1529} & 
\begin{tabular}[c]{@{}r@{}}1430~\cite{norton1973impurity}\\ \textbf{1500}~\cite{sze2008semiconductor}\end{tabular} \\[20pt]

Diamond & 1290 & 
\begin{tabular}[c]{@{}r@{}}1800~\cite{haynes2020crc}\\
1802~\cite{jansen2013temperature}\\ 1940~\cite{gabrysch2011electron}\\\end{tabular} 
& \textbf{1172} & 
\begin{tabular}[c]{@{}r@{}} \textbf{1030}~\cite{pernot2006hall} \\ 1800~\cite{redfield1954electronic} \end{tabular} \\[20pt]
GaAs & 9680 & 
\begin{tabular}[c]{@{}r@{}}7390~\cite{STILLMAN19701199}\end{tabular} 
& \textbf{10424} & 
\begin{tabular}[c]{@{}r@{}}\\ 8500~\cite{rode1971electron}\\ 8725~\cite{STILLMAN19701199}\\ \textbf{8900}~\cite{hicks1969high}\end{tabular} \\[3pt]
\hline\\[-9.5pt] \hline\\[-7pt]
\end{tabular}
\label{tab:mobility_comparison}
\end{table}

\vspace*{\fill}

\clearpage
\newpage

\vspace*{\fill}

\begin{table} \color{black}
\caption{\textbf{Cross-validation for mobility error statistics.} Jackknife resampling of the mobility error compared to experimental measurements, at various levels of theory. The error is recomputed by systematically excluding one material to check the sensitivity of the average error to the chosen sample. The mobilities of Fig.~2(a) are used for this statistical analysis.}
\vspace{10pt}
\centering
\begin{tabular}{@{\extracolsep{\fill}} l @{\hspace{0.5cm}} r @{\hspace{0.5cm}} r  @{\hspace{0.5cm}} r}
\hline\\[-9.5pt] \hline\\[-5pt]
Excluded Sample & \multicolumn{3}{c}{Average Error} \\[4pt]
                  & DFT+DFPT & GW+DFPT & GW+GWPT \\
                  & \% & \% & \% \\[2pt]
&  & \\[-8pt]
\hline\\[-5pt]
GaP      & 130 & 37.4 & 12.6\\
3C-SiC   & 120 & 37.4 & 12.6\\
Si       & 130 & 52.2 & 13.4\\
Diamond  & 126 & 42.3 & 10.4\\
GaAs     & 24.8 & 42.2 & 12.6\\
None     & 106 & 42.1 & 11.1 \\[3pt]
\hline\\[-9.5pt] \hline\\[-7pt]
\end{tabular}
\end{table}

\clearpage 
\begin{longtable}{@{\extracolsep{\fill}} l c r r r c}
\caption{\textbf{Calculated and experimental mobilities reported in Fig.~2(c).} 
Details of mobility data presented in Fig.~2(c), including temperature, measurement type, and data provenance. Experimental valued were extracted using a graph digitizer~\cite{WebPlotDigitizer} unless explicitly given in the cited article. The value marked by the dagger ($\dagger$) includes a contact-size correction of +30\% determined in the experimental paper. We report the calculated drift or Hall mobility according to the experiment type in the fifth column.\vspace{10pt}}\\
\hline\\[-9.5pt] \hline\\[-5pt]
Material & Temperature & Theory & Experiment & Measurement & Reference \\
 & (K) & (cm$^2$/Vs) & (cm$^2$/Vs) & & \\[2pt]
\hline\\[-5pt]
\endfirsthead

\multicolumn{6}{c}{{ \tablename\ \thetable{} continued}} \\[3pt]
\hline\\[-9.5pt] \hline\\[-5pt]
Material & Temperature & Theory & Experiment & Measurement & Reference \\
 & (K) & (cm$^2$/Vs) & (cm$^2$/Vs) & & \\[2pt]
\hline\\[-5pt]
\endhead

\hline\\[-9.5pt] \hline\\[-7pt]
\multicolumn{6}{r}{{Continued on next page}} \\
\endfoot

\hline\\[-9.5pt] \hline\\[-7pt]
\endlastfoot

GaAs & 279 & 11732 & 12283 & Hall & \cite{hicks1969high} \\
GaAs & 316 & 9628 & 9385 & Hall & \cite{hicks1969high} \\
GaAs & 339 & 8656 & 8327 & Hall & \cite{hicks1969high} \\
GaAs & 396 & 6994 & 5821 & Hall & \cite{hicks1969high} \\
GaAs & 259 & 13287 & 13735 & Hall & \cite{rode1971electron} \\
GaAs & 319 & 9470 & 8433 & Hall & \cite{rode1971electron} \\
GaAs & 277 & 11845 & 10308 & Hall & \cite{stanley19914} \\
GaAs & 252 & 13866 & 17739 & Hall & \cite{lin1982vapour} \\
GaAs & 266 & 12659 & 15569 & Hall & \cite{lin1982vapour} \\
GaAs & 296 & 10651 & 11992 & Hall & \cite{lin1982vapour} \\
GaAs & 307 & 10070 & 10057 & Hall & \cite{lin1982vapour} \\
GaAs & 300 & 10424 & 8865 & Hall  & \cite{madelung2004semiconductors} \\
GaAs & 300 & 10424 & 8500 & Hall & \cite{rode1971electron} \\
GaAs & 300 & 10424 & 8725 & Hall & \cite{STILLMAN19701199} \\
GaAs & 300 & 10424 & 8900 & Hall & \cite{hicks1969high} \\
GaAs & 300 & 10424 & 9000 & Hall & \cite{madelung2004semiconductors} \\
GaAs & 300 & 9680 & 7390 & Drift & \cite{STILLMAN19701199} \\
C & 251 & 1733 & 3155 & Drift & \cite{gabrysch2011electron} \\
C & 274 & 1507 & 2635 & Drift & \cite{gabrysch2011electron} \\
C & 296 & 1305 & 2281 & Drift & \cite{gabrysch2011electron} \\
C & 334 & 914 & 1710 & Drift & \cite{gabrysch2011electron} \\
C & 352 & 830 & 1449 & Drift & \cite{gabrysch2011electron} \\
C & 367 & 748 & 1291 & Drift & \cite{gabrysch2011electron} \\
C & 385 & 748 & 1150 & Drift & \cite{gabrysch2011electron} \\
C & 400 & 683 & 1033 & Drift & \cite{gabrysch2011electron} \\
C & 413 & 634 & 913 & Drift & \cite{gabrysch2011electron} \\
C & 427 & 586 & 802 & Drift & \cite{gabrysch2011electron} \\
C & 439 & 550 & 741 & Drift & \cite{gabrysch2011electron} \\
C & 315 & 1150 & 2454 & Drift & \cite{nava1980electron} \\
C & 402 & 675 & 1510 & Drift & \cite{nava1980electron} \\
C & 253 & 1612 & 1781 & Hall & \cite{redfield1954electronic} \\
C & 293 & 1225 & 1451 & Hall & \cite{redfield1954electronic} \\
C & 300 & 1290 & 1802 & Drift & \cite{jansen2013temperature} \\
C & 300 & 1290 & 1940 & Drift & \cite{gabrysch2011electron} \\
C & 300 & 1290 & 2000 & Drift & \cite{madelung2004semiconductors} \\
C & 300 & 1290 & 2750 & Drift & \cite{nesladek2008charge} \\
SiC & 449 & 449 & 322 & Hall & \cite{nelson1966growth} \\
SiC & 380 & 700 & 581 & Hall & \cite{nelson1966growth} \\
SiC & 300 & 1379 & 1216 & Hall & \cite{nelson1966growth} \\
SiC & 447 & 453 & 313 & Hall & \cite{shinohara1988growth} \\
SiC & 419 & 538 & 398 & Hall & \cite{shinohara1988growth} \\
SiC & 383 & 688 & 539 & Hall & \cite{shinohara1988growth} \\
SiC & 322 & 1126 & 763 & Hall & \cite{shinohara1988growth} \\
SiC & 291 & 1510 & 949 & Hall & \cite{shinohara1988growth} \\
SiC & 300 & 1377 & 890 & Hall & \cite{nelson1966growth} \\
SiC & 300 & 1377 & 980 & Hall & \cite{nelson1966growth} \\
SiC & 300 & 1377 & 1000 & Hall & \cite{bhatnagar1993ieee} \\
SiC & 300 & 1377 & 1157$^\dagger$ & Hall & \cite{nelson1966growth} \\
Si & 280 & 1754 & 2213 & Hall & \cite{norton1973impurity} \\
Si & 366 & 1039 & 1027 & Hall & \cite{norton1973impurity} \\
Si & 326 & 1300 & 1457 & Hall & \cite{norton1973impurity} \\
Si & 318 & 1366 & 1275 & Hall & \cite{jacoboni1977review} \\
Si & 386 & 936 & 848 & Hall & \cite{jacoboni1977review} \\
Si & 422 & 788 & 576 & Hall & \cite{jacoboni1977review} \\
Si & 300 & 1529 & 1430 & Hall & \cite{norton1973impurity} \\
Si & 300 & 1529 & 1500 & Hall & \cite{sze2008semiconductor} \\
Si & 300 & 1390 & 1350 & Drift & \cite{madelung2004semiconductors} \\
Si & 300 & 1390 & 1350 & Drift & \cite{ludwig1956drift} \\
Si & 300 & 1390 & 1350 & Drift & \cite{norton1973impurity} \\
Si & 300 & 1390 & 1800 & Drift & \cite{haynes2020crc} \\
GaP & 356 & 222 & 196 & Hall & \cite{miyauchi1967electrical} \\
GaP & 337 & 248 & 218 & Hall & \cite{miyauchi1967electrical} \\
GaP & 301 & 312 & 256 & Hall & \cite{miyauchi1967electrical} \\
GaP & 264 & 413 & 296 & Hall & \cite{miyauchi1967electrical} \\
GaP & 300 & 315 & 256 & Hall & \cite{miyauchi1967electrical} \\
GaP & 300 & 315 & 300 & Hall & \cite{haynes2020crc} \\

\end{longtable}

\vspace*{\fill}

\clearpage
\newpage

\begin{table}
\caption{\textbf{Comparison between the \textit{ai}BTE and SERTA mobilities.} Room-temperature Drift mobilities calculated with GWBTE. The SERTA approximation [Eq.~\eqref{eq:serta}] captures the trends of the full \textit{ai}BTE results [Eq.~\eqref{eq:aibte}], therefore it can meaningfully be used to perform the Drude-like decomposition in Eq.~\eqref{eq:factorized}.}
\vspace{4pt}
\centering
\begin{tabular}{@{\extracolsep{\fill}} l @{\hspace{0.5cm}}r @{\hspace{0.5cm}}r }
\hline\\[-9.5pt] \hline\\[-5pt]
Material & SERTA & \textit{ai}BTE \\
 & (cm$^2$/Vs) & (cm$^2$/Vs) \\[2pt]
&  \\[-8pt]
\hline\\[-5pt]
GaP     & 216  & 271 \\
Diamond & 1149 & 1290  \\
3C-SiC      & 1227 & 1265 \\
Si       & 1350 & 1390 \\
GaAs    & 5924 & 9680 \\[3pt]
\hline\\[-9.5pt] \hline\\[-7pt]
\end{tabular}
\end{table}

\vspace*{\fill}

\clearpage
\newpage
\begin{table}[htp!] \color{black}
\caption{\textbf{Full-frequency calculation of the GW electron--phonon interaction.} Comparison of the supercell GW electron--phonon interaction calculated within the generalized plasmon pole approximation versus the full-frequency calculation with contour deformation in diamond. The three-fold splitting of the conduction band at the $\Gamma$-point is shown due to a frozen phonon corresponding to the $\Gamma$-point displacement pattern of the optical phonon mode of diamond in the unit cell.}
    \vspace{4pt}
    \centering
    \begin{tabular}{@{\extracolsep{\fill}}  r @{\hspace{0.5cm}} r @{\hspace{0.5cm}}}
    \hline\\[-9.5pt] \hline\\[-5pt]
    Plasmon-Pole Model & Full Frequency \\
    (eV) & (eV) \\
    &  \\[-8pt]
    \hline\\[-5pt]
     $-$0.743 & $-$0.744\\
     0.356 & 0.366 \\
     0.382 & 0.392\\[3pt]
    \hline\\[-9.5pt] \hline\\[-7pt]
    \end{tabular}
\end{table}

\clearpage
\newpage

\begin{table}
\caption{\textbf{Comparison between calculated and experimental band gaps.} We report our $G_0W_0$ calculations of the band gaps of the materials considered in this work, and the corresponding experimental data from Ref.~\citenum{madelung2004semiconductors}.}
\vspace{4pt}
\centering
\begin{tabular}{@{\extracolsep{\fill}} l @{\hspace{0.5cm}} r @{\hspace{0.8cm}} r @{\hspace{0.5cm}} c}
\hline\\[-9.5pt] \hline\\[-5pt]
 & \multicolumn{2}{c}{\hspace{-10pt}Theory} & Experiment \\ 
\cmidrule(rr){2-3}
Material & DFT  & GW  & \\
& (eV) & (eV) & (eV) \\[2pt]
&  \\[-8pt]
\hline\\[-5pt]
GaP     & 1.38 & 2.52 & 2.35\\
Diamond & 4.18 & 5.61 & 5.51 \\
Si      & 0.61 & 1.29 & 1.20 \\
3C-SiC  & 1.40 & 2.48 & 2.42\\
GaAs    & 0.49 & 1.22 & 1.52 \\[3pt]
\hline\\[-9.5pt] \hline\\[-7pt]
\end{tabular}
\end{table}

\clearpage
\newpage

\begin{table} \color{black}
\caption{\textbf{Convergence of GW calculations in diamond.} Calculation denotes the specific calculation for which the convergence test is performed.
Quantity refers to the physical quantity being tested for convergence.
Band energy corresponds to the band cutoff, defined as the highest energy included in the number of bands used in the summation over empty states, referenced to the valence band top.}
\vspace{4pt}
\centering
\begin{tabular}{@{\extracolsep{\fill}} l @{\hspace{1cm}} l @{\hspace{1.5cm}} r @{\hspace{0.5cm}} r @{\hspace{0.5cm}} r @{\hspace{0.5cm}} r @{\hspace{0.5cm}} r}
\hline\\[-9.5pt] \hline\\[-5pt]
& &  \multicolumn{5}{c}{\hspace{-10pt} Band energy cutoff (eV)} \\ 
Calculation & Quantity & 90.9 & 210 & 263 & 316 & 353 \\[2pt]
\\[-8pt]
\hline\\[-5pt]
\multirow{2}{*}{Dielectric function (epsilon)} & $\text{E}_\text{gap}^\Gamma$  &  &  7.46 & 7.47 & 7.47 & 7.47\\
                         & $m^*_\text{cond.}$ &  & 0.391 & 0.390 & 0.390 & 0.390 \\[5pt]
\multirow{2}{*}{Self-energy (sigma)} & $\text{E}_\text{gap}^\Gamma$  &  7.48 &  7.47 & 7.47 & 7.47 & 7.47\\
                         & $m^*_\text{cond.}$ & 0.393 & 0.391 & 0.390 & 0.390 & 0.390 \\[5pt]
\hline\\[-9.5pt] \hline\\[-7pt]
\end{tabular}
\end{table}

\clearpage
\newpage

\begin{table} \color{black}
\caption{\textbf{Convergence of GWPT calculations in diamond.} The convergence test shown is for the expectation value, $\bra{n\textbf{k}} \Delta_{\textbf{q}\kappa\alpha} \Sigma - \Delta_{\textbf{q}\kappa\alpha} V^\text{XC} \ket{n\textbf{k}}$, for arbitrarily chosen electron--phonon matrix elements and for the summation over empty states in constructing the variation of the GW self-energy. Band energy corresponds to the band cutoff, defined as the highest energy included in the number of bands used in the summation over empty states, referenced to the valence band top.}
\vspace{4pt}
\centering
\begin{tabular}{@{\extracolsep{\fill}} c @{\hspace{0.5cm}} c @{\hspace{0.5cm}}  l @{\hspace{0.5cm}} l @{\hspace{0.5cm}} r @{\hspace{0.5cm}} r @{\hspace{0.5cm}} r @{\hspace{0.5cm}} r @{\hspace{0.5cm}} r}
\hline\\[-9.5pt] \hline\\[-5pt]
& & & & \multicolumn{5}{c}{\hspace{-10pt} Band energy cutoff (eV)} \\ 
q & k & $\kappa$ & $\alpha$ & 175 & 263 & 353 & 387 & 435 \\[2pt]
\\[-8pt]
\hline\\[-5pt]
(0, 0, 0) & (0, 0, 0)  & 1 & $x$ & $-$0.003 & $-$0.006 & $-$0.008 & $-$0.008 & $-$0.009\\
(0, 0, 0) & (0, 0, 1/8) & 1 & $x$ & $-$3.58 & $-$3.55 & $-$3.55 & $-$3.54 & $-$3.54 \\
(0, 0, 0) & (0, 0, 1/4) & 1 & $x$ & $-$3.89 & $-$3.85 & $-$3.85 & $-$3.84 & $-$3.84 \\
(1/2, 0, 0) & (0, 0, 0) & 1 & $z$ & $-6.96 + i0.03$ & $-6.90 + i0.03$ & $-6.90 + i0.03$ & $-6.90 + i0.03$ & $-6.90 + i0.03$ \\
(1/2, 0, 0) & (1/4, 0, 0) & 1 & $z$ & $-6.42 - i2.49$ & $-6.37 - i2.48$ & $-6.36 - i2.48$ & $-6.36 - i2.48$ & $-6.36 - i2.48$ \\
(1/2, 0, 0) & (3/8, 1/8, 5/8) & 1 & $z$  &  $-1.06 - i0.43$ & $-1.02 - i0.42$ & $-1.01 - i0.41$ & $-1.00 - i0.41$ & $-1.00 - i0.41$  \\
(1/4, 1/4, 0) & (0, 0, 0) & 2 & $x$ &  $-0.39 + i1.73$ & $-0.39 + i1.72$ & $-0.39 + i1.71$ & $-0.39 + i1.71$ & $-0.39 + i1.71$\\
(1/4, 1/4, 0) & (7/8, 1/4, 3/4) & 2 & $x$ &  $-4.60 + i1.20$ & $-4.55 + i1.16$ & $-4.53 + i1.13$ & $-4.53 + i1.13$ & $-4.52 + i1.12$ \\
(1/4, 1/4, 0) & (7/8, 7/8, 7/8) & 2 & $x$ &  $-1.12 + i0.86$ & $-1.12 + i0.85$ & $-1.12 + i0.85$ & $-1.12 + i0.85$ & $-1.12 + i0.85$\\[5pt]
\hline\\[-9.5pt] \hline\\[-7pt]
\end{tabular}
\end{table}

\clearpage
\newpage

\begin{table} \color{black}
\caption{\textbf{Computational cost of GWPT calculations.} Calculations were performed on the CPU nodes of the Frontera supercomputer at the Texas Advanced Computing Center. N$_\textbf{k}$ and N$_\textbf{q}$ refer to the number of k and q points used in the GWPT calculation, and N$_\text{bands}$ refers to the number of conduction bands for which the matrix elements are corrected.}
\vspace{4pt}
\centering
\begin{tabular}{@{\extracolsep{\fill}} l @{\hspace{0.5cm}} r @{\hspace{0.5cm}} r @{\hspace{0.5cm}} r @{\hspace{0.5cm}} r}
\hline\\[-9.5pt] \hline\\[-5pt]
Material & N$_\textbf{k}$ & N$_\textbf{q}$ & N$_\text{bands}$ & Cost \\
& & & & core hours\\[2pt]
&  \\[-8pt]
\hline\\[-5pt]
GaP     & 512 & 64 & 2 &  532,141\\
Diamond & 512 & 512 & 1 & 1,596,336 \\
Si      & 512 & 64 & 2 & 553,595\\
3C-SiC  & 512 & 64 & 1 & 647,548\\
GaAs    & 64 & 64 & 1 & 57,524\\
\textbf{Total} & & & & \textbf{3,387,144} \\[3pt]
\hline\\[-9.5pt] \hline\\[-7pt]
\end{tabular}
\end{table}

\clearpage
\begin{figure}
    \centering
    \includegraphics[width=0.8\linewidth]{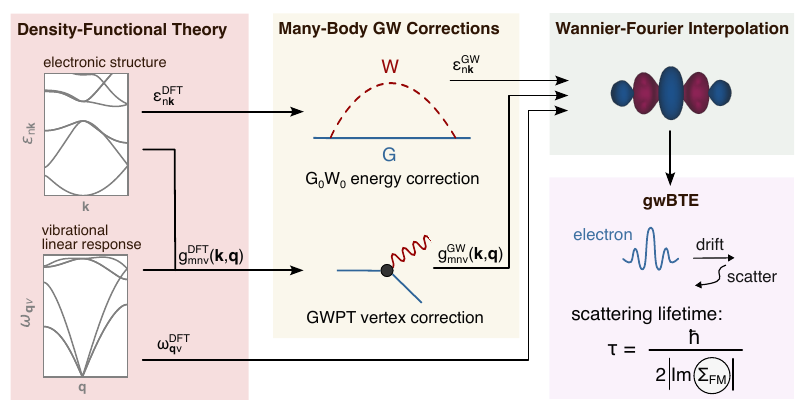}
    \caption{\textbf{Schematic workflow of GWBTE calculation.} Calculation steps include DFT and DFPT calculations of electron bands and phonons, respectively; GW calculations of band renormalization; GWPT calculations of electron-phonon coupling corrections; Wannier interpolation of all quantities to fine Brillouin-zone grids; evaluation of scattering rates; and solution of the \textit{ai}BTE.}
    \label{fig:workflow}
\end{figure}

\begin{figure}
    \centering
    \includegraphics[width=0.6\linewidth]{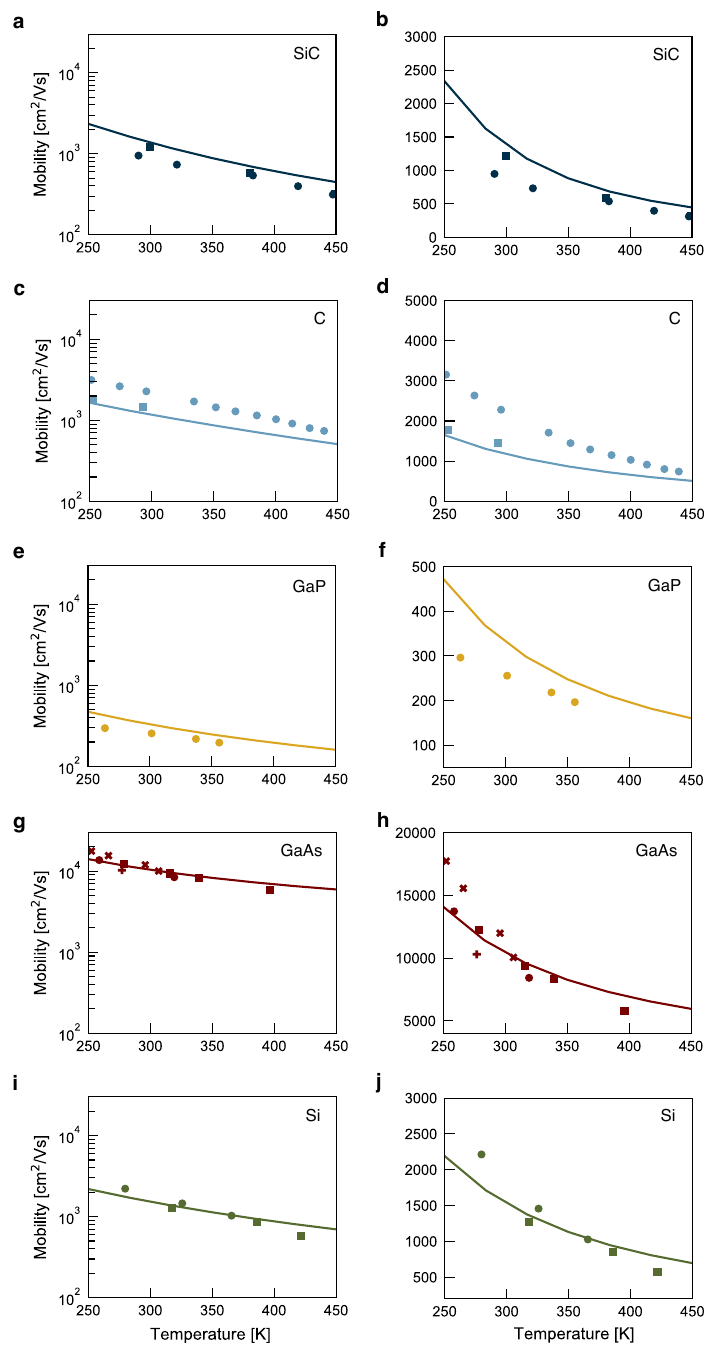}
    \caption{\textbf{Temperature-dependent Hall mobilities.} Comparison between our calculated GWBTE mobilities and experiments, as a function of temperature in the range 250-450\,K. Panels (a), (c), (e), (g), and (i) show the data in logarithmic scale, while panels (b), (d), (f), (h), and (j) show the same data in linear scale. The lines are our calculations. The data points are experimental data from the following references:
    3C-SiC: circles~\cite{shinohara1988growth}, squares~\cite{nelson1966growth}; 
    diamond: circles~\cite{redfield1954electronic}, squares~\cite{gabrysch2011electron}; 
    GaP: circles~\cite{miyauchi1967electrical};
    GaAs: circles~\cite{rode1971electron}, squares~\cite{hicks1969high}, pluses~\cite{stanley19914}, crosses~\cite{lin1982vapour};
    Si: circles~\cite{jacoboni1977review}, squares~\cite{norton1973impurity}. 
    Thermal expansion is not taken into account in the calculations.}
    \label{fig:temp_dep}
\end{figure}


\vspace*{\fill}
\begin{figure}
    \centering
    \includegraphics[width=0.45\linewidth]{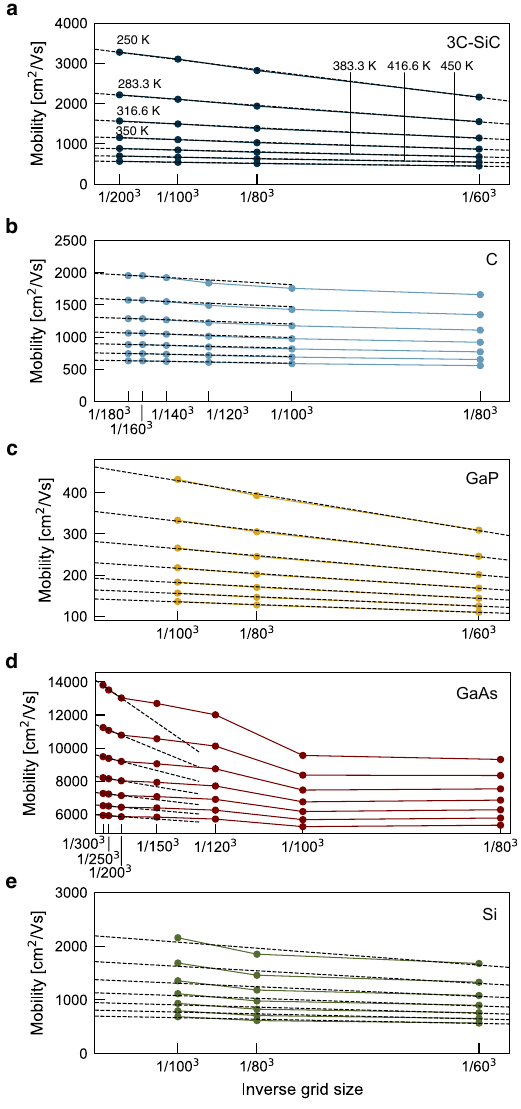}
    \caption{\textbf{Convergence tests for the calculated mobilities.}
    Numerical convergence of the temperature-dependent mobilities shown in Supplemental Fig.~2: (a) 3C-SiC, (b) diamond, (c) GaP, (d) GaAs, and (e) Si.
    The lines connect datasets at the same temperature, for seven temperature values equally spaced bwteen 250\,K and 450\,K. The horizontal axis is the size of the fine Brillouin zone grid used in mobility calculations; for example, $1/300^3$ means a grid with 300$\times$300$\times$300 $\textbf{k}$- and $\textbf{q}$-points.    
    The finest grids employed correspond to $100^3$ points for GaP, $200^3$ points for 3C-SiC, $100^3$ points for Si, $180^3$ points for diamond, and $300^3$ points for GaAs. All mobility values reported in the main text were obtained from extrapolation to the infinite sampling limit ($N\!\times\! N\!\times\! N$ points with $N \to \infty$) using a linear fit in $1/N^3$, as shown by the dashed lines.}
\end{figure}
\vspace*{\fill}

\clearpage

\begin{figure}
    \centering
    \includegraphics[width=0.8\linewidth]{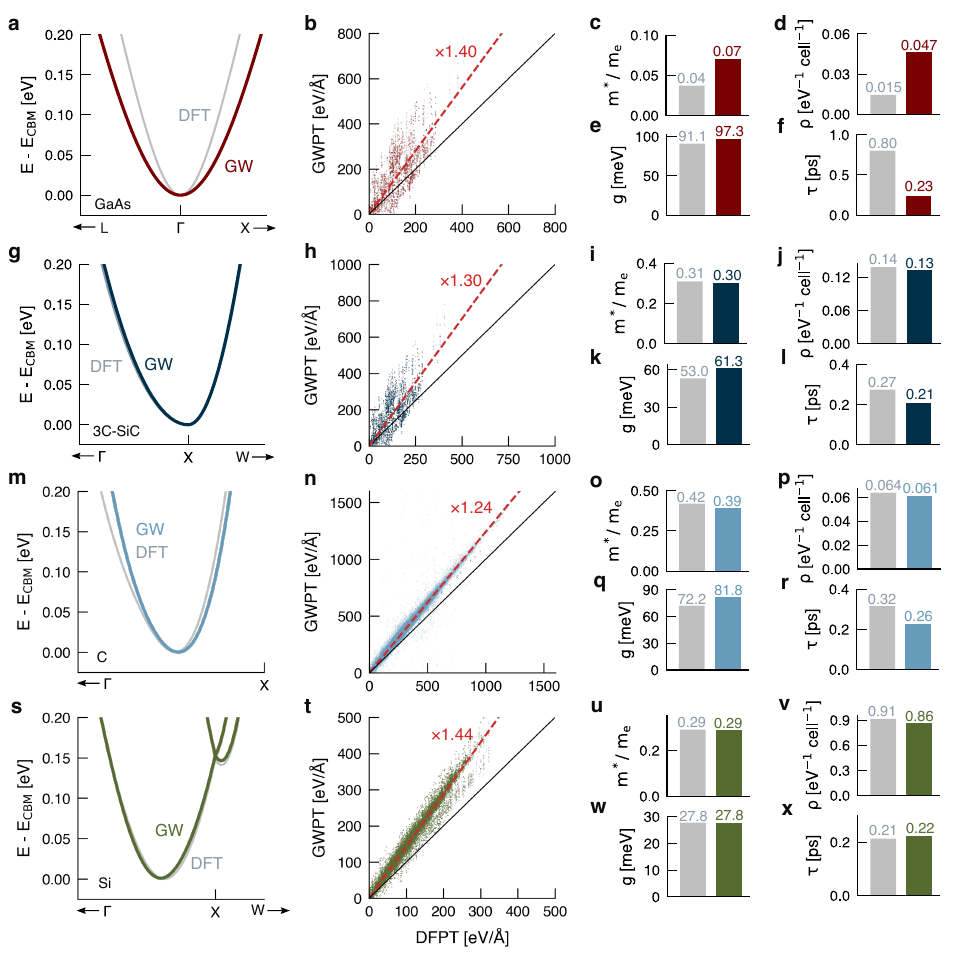}
    \caption{\textbf{Disentangling many-body contributions to the mobility.} 
    (a) Comparison between DFT (gray) and GW (red) band structures of GaAs.
    (b) Comparison between DFPT and GWPT electron-phonon matrix elements for GaAs. The solid line indicates a 1:1 ratio, the dashed line is the average ratio over the entire dataset. The matrix element is given in the displacement representation:
     $|\bra{\psi_{m\textbf{k}+\textbf{q}}} \Delta_{\kappa\alpha,\textbf{q}} V_{KS} \ket{\psi_{n\textbf{k}}}|$, and is evaluated on a uniform coarse Brillouin zone grid. (c) Comparison between the effective DFT (gray) and GWPT (red) electron mass of GaAs, as defined by Eq.~\eqref{eq:crta}.  (d) Comparison between the DFT (gray) and GW (red) effective scattering density of states in GaAs, as calculated from Eq.~\eqref{eq:rho}. (e) Comparison between the DFT (gray) and GW (red) effective electron-phonon coupling $g$ in GaAs, as defined by Eq.~\eqref{eq:g2}. The average increase of electron-phonon coupling for states across the Brillouin zone (40\% in (b)) is generally not equivalent to the increase of $g$ relevant for band-edge states that participate in transport (6.8\% in (e)). (f)  Comparison between the DFT (gray) and GW (red) effective electron lifetimes, as obtained from Eq.~\eqref{eq:crta2}. (j)-(l): Same information as in (a)-(f), but for 3C-SiC.
     (m)-(r): Same information as in (a)-(f), but for diamond.
    (s)-(x): Same information as in (a)-(f), but for Si.
     }
\end{figure}

\clearpage
\begin{figure}
    \centering
    \includegraphics[width=0.55\linewidth]{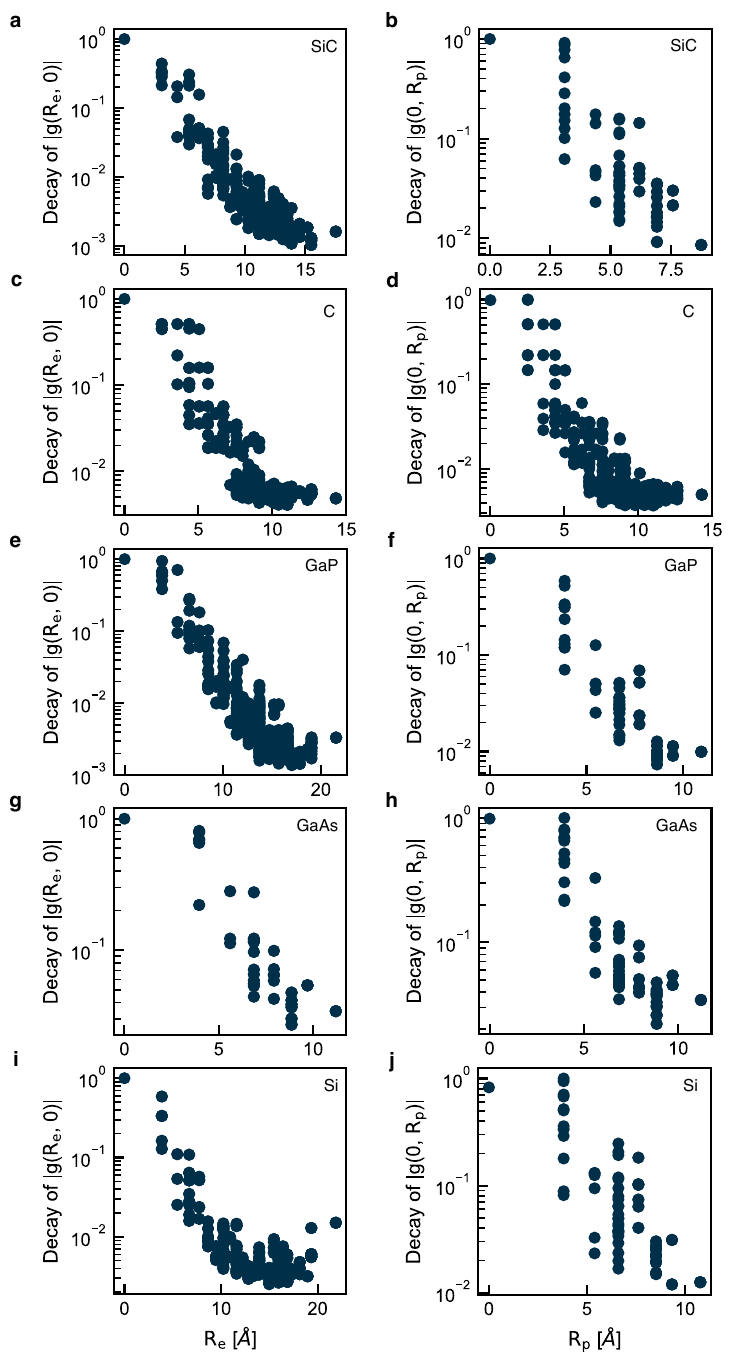}
    \caption{\textbf{Spatial localization of the GWPT electron-phonon matrix elements.} (a) Dependence of the electron-phonon matrix elements in the Wannier representation, $g^{\rm GW}_{mn,\kappa \alpha}(\textbf{R}_{\rm e},\textbf{R}_{\rm p})$, on the separation $|\textbf{R}_{\rm e}|$ between Wannier functions, for 3C-SiC. We report the modulus of the matrix element for every $m,n,\kappa\alpha$. These data correspond to the short-range part of the matrix elements, after removing long-range dipole and quadrupole terms~\cite{verdi2015frohlich,brunin2020electron}.
    (b) Same as in (a), except that the matrix elements correspond to two Wannier functions at $\textbf{R}_{\rm e}=0$, and we vary the location of the atomic dipole via $\textbf{R}_{\rm p}$. 
    (c)-(d) Same as in (a)-(b), but for diamond.
    (e)-(f) Same as in (a)-(b), but for GaP.
    (g)-(h) Same as in (a)-(b), but for GaAs.
    (i)-(j) Same as in (a)-(b), but for Si.
    }
    \label{fig:decay}
\end{figure}

\clearpage
\newpage

\begin{figure} \color{black}
    \centering
    \includegraphics[width=0.6\linewidth]{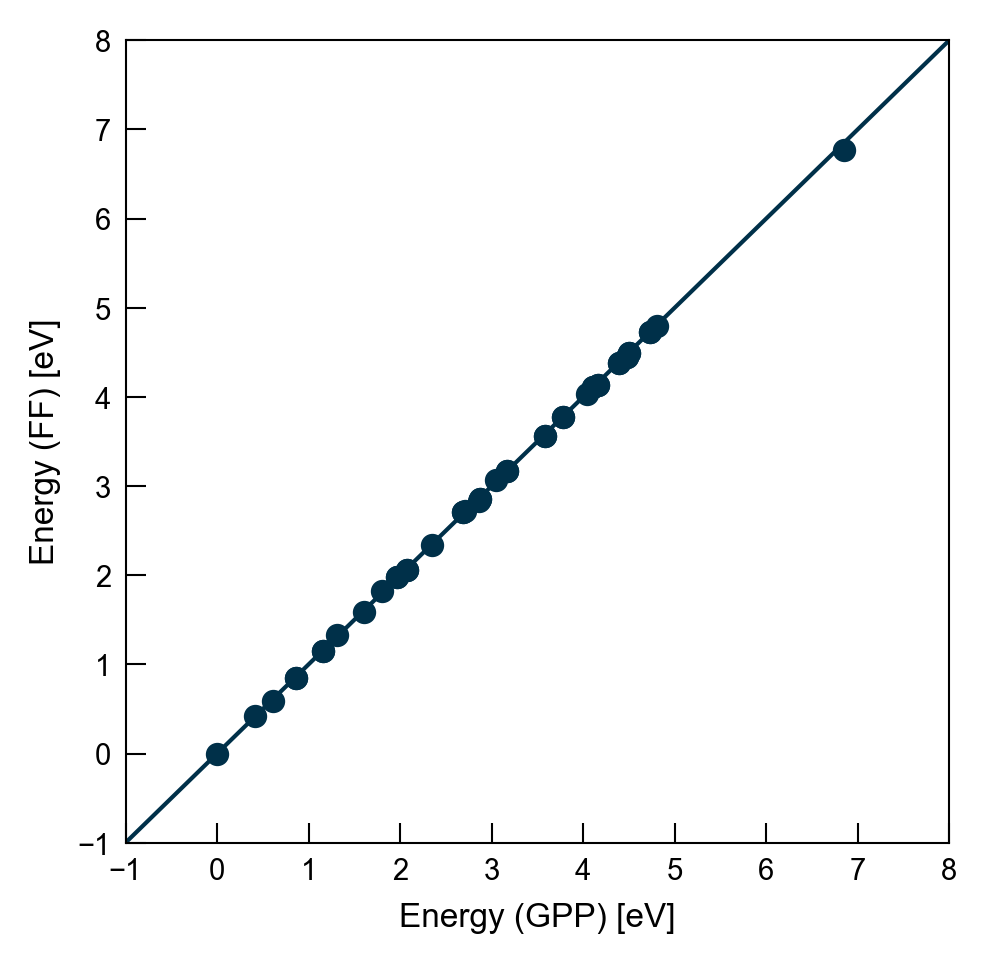}
    \caption{\textbf{Full-frequency calculation of GW energies.} Comparison of the GW energy calculated within the generalized plasmon pole (GPP) approximation to the full-frequency (FF) calculation with contour deformation in diamond. The energies corresponding to states in the lowest conduction band of diamond are shown, referenced to the conduction-band bottom. The mean absolute error of GW(GPP) vs GW(DFT) is 0.014 eV; in contrast, the mean absolute error of DFT versus GW(FF) is 0.28 eV.}
\end{figure}
\clearpage
\newpage

\bibliographystyle{apsrev4-2}
\bibliography{supp_bib}